\documentclass[iop]{emulateapj} 
\usepackage{times}
\usepackage{natbib}
\usepackage{rotating}
\newcommand{\oiii}{{[O\,{\sc iii}]}}

\newcommand{\nii}{{[N\,{\sc ii}]}}

\newcommand{\hb}{{H$\beta$}}
\newcommand{\ha}{{H$\alpha$}}

\newcommand{\kms}{\rm km~s\ensuremath{^{-1}\,}}
\newcommand{\fourstar}{{\sc FourStar}}

\begin{document}
\title{KECK/MOSFIRE spectroscopic confirmation\altaffilmark{1}  of a Virgo-like cluster ancestor at z{=}2.095}

\author{Tiantian~Yuan\altaffilmark{2},
Themiya~Nanayakkara\altaffilmark{3},
Glenn~G.~Kacprzak\altaffilmark{3,4},
Kim-Vy~H.~Tran\altaffilmark{5},
Karl~Glazebrook\altaffilmark{3},
Lisa J. Kewley\altaffilmark{2},
Lee~R.~Spitler\altaffilmark{6,7},
Gregory~B.~Poole\altaffilmark{8}, 
Ivo~Labb\'e\altaffilmark{9},
Caroline M. S.~Straatman\altaffilmark{9},
Adam R.~Tomczak\altaffilmark{5}
}

\altaffiltext{1}{Based on data obtained at the
W.M. Keck Observatory, which
is operated as a scientific partnership among the California Institute of
Technology, the University of California, and NASA, and was made possible by the generous
financial support of the W.M. Keck Foundation.
}
\altaffiltext{2}{Research School of Astronomy and Astrophysics, The Australian National University, Cotter Road, Weston Creek, ACT 2611}
\altaffiltext{3}{Centre for Astrophysics \& Supercomputing, Swinburne University, Hawthorn, VIC 3122, Australia}
\altaffiltext{4}{Australian Research Council Super Science Fellow}
\altaffiltext{5}{George P.\ and Cynthia Woods Mitchell Institute for Fundamental  Physics and Astronomy, and Department of Physics and Astronomy, Texas A\&M University, College Station, TX, 77843-4242, USA}
\altaffiltext{6}{Department of Physics \& Astronomy, Macquarie University, Sydney, NSW 2109, Australia}
\altaffiltext{7}{Australian Astronomical Observatory, P.O. Box 296 Epping, NSW 1710, Australia}
\altaffiltext{8}{School of Physics, University of Melbourne, Parksville, VIC 3010, Australia}
\altaffiltext{9}{Sterrewacht Leiden, Leiden University, NL-2300 RA Leiden, The Netherlands}

\begin{abstract}

We present the spectroscopic confirmation of a  galaxy cluster at $z{=}2.095$ in the COSMOS field.
This galaxy cluster was first reported in the ZFOURGE survey as harboring evolved massive galaxies using photometric redshifts derived with deep
near-infrared (NIR) medium-band filters.  We obtain medium resolution ($R \sim$ 3600)
NIR spectroscopy with MOSFIRE on the Keck 1 telescope and secure 180 redshifts  in a $12'\times12'$ region. 
We find a prominent spike of 57 galaxies at $z{=}2.095$ corresponding to the galaxy cluster. 
The cluster velocity dispersion is measured to be $\sigma_{\rm v1D}$ {=} 552 $\pm$ 52 km/s. 
This is the first study of a galaxy cluster in this redshift range ($z \ga 2.0$) with the combination
of spectral resolution ($\sim$ 26 km/s) and the number of confirmed members (${>}50$) needed to 
impose a meaningful constraint on the cluster velocity dispersion and map its members over a large field of view.
Our $\Lambda$CDM cosmological simulation suggests that this cluster will most likely evolve into a  Virgo-like 
 cluster with ${\rm M_{vir}}{=}10^{14.4\pm0.3} {\rm M_\odot}$ ($68\%$ confidence) at $z
\sim$ 0.    The theoretical expectation of finding such a cluster is $\sim$ 4\%. 
Our results demonstrate the feasibility of studying galaxy clusters at $z > 2$ in the same detailed manner using multi-object NIR spectrographs  as has been done in the optical in lower redshift clusters.
\end{abstract}


\section{Introduction}

In the standard cosmological model of structure formation, galaxy
clusters are the largest collapsing structures located at the nodes of
the cosmic web.  Studies of local galaxies have found strong
correlations of galaxy properties with the environment
\citep[e.g.,][]{Dressler80,Hogg04}.  However, it is largely
unknown whether and how these correlations would hold up at higher
redshifts of $z\ga2$, when the mean star formation activities of the
universe peaked and clusters were formed
\citep[e.g.,][]{HopkinsAM06,Rettura10}.  Studying dense galaxy groups
and clusters at $z\ga2$ provides crucial knowledge of the star
formation history of high-mass galaxies and the hierarchical growth of
massive structures
\citep[e.g.,][]{McCarthy07,Tran10,Brodwin13,Strazzullo13,Henry14}.

Great progress has been made in increasing the number of cluster
candidates at $z\ga 1.6$ \citep[e.g.,][]{Papovich10,Hayashi12,Muzzin13,LeeK14,
Chiang14,Newman14}.  To secure the identification of a cluster, and to
further elucidate the star formation history and physical properties
of the galaxy members, spectroscopic follow-up is necessary.  Because
of the amount of large telescope time required, it is not surprising
that to date only a handful of spectroscopically-confirmed galaxy
clusters with developed red sequences are known at $z\ga 2$.  Existing
studies either do not have accurate cluster velocity dispersion
measurements due to small numbers (${<}10$) resulting in large
uncertainties \citep{Kurk04,Galametz13}, or membership comes from
Hubble Space Telescope (HST) grism redshifts with typical redshift
accuracies of $\pm$ 200 km/s on individual galaxies
\citep{Gobat13}. Non-uniform redshift identifications from different
instruments with limited spectral resolution and sensitivity also
makes it difficult to quantify the errors of cluster velocity dispersion
\citep{Shimakawa14}.

We capitalize on the efficient Multi-Object Spectrometer for InfraRed
Exploration (MOSFIRE; \citealt{McLean10,McLean12}) on KECK-1 to carry
out a uniform spectroscopic survey on a galaxy cluster at $z\sim 2$
\citep{Spitler12} which was first identified using deep medium-band
photometry in the \fourstar\ \citep{Persson13} Galaxy Evolution Survey
(ZFOURGE\footnote{{\url http://zfourge.tamu.edu}}) as having a
striking overdensity in red galaxies.  In this Letter we use MOSFIRE
to spectroscopically confirm 57 cluster members (spectral resolution
$\sim$ 10 km/s) and accurately measure the galaxy cluster's velocity
dispersion.  Our study also confirms the robustness of the ZFOURGE
photometric redshifts and ability to detect galaxies at $z\ga 2$.

Throughout the paper, we adopt a flat cosmology with
$\Omega_{M}{=}0.3$, $\Omega_{\Lambda}{=}0.7$ and
H$_{0}{=}70$~km~s$^{-1}$~Mpc$^{-1}$.  At the cluster redshift of
$z{=}2.09$, 10 arcmin corresponds to an angular scale of 5 Mpc in
proper coordinates.

\begin{figure*}[!ht]
\centering
\includegraphics[trim = 6mm 0mm 2mm 0mm, clip, width=18cm,angle=0]{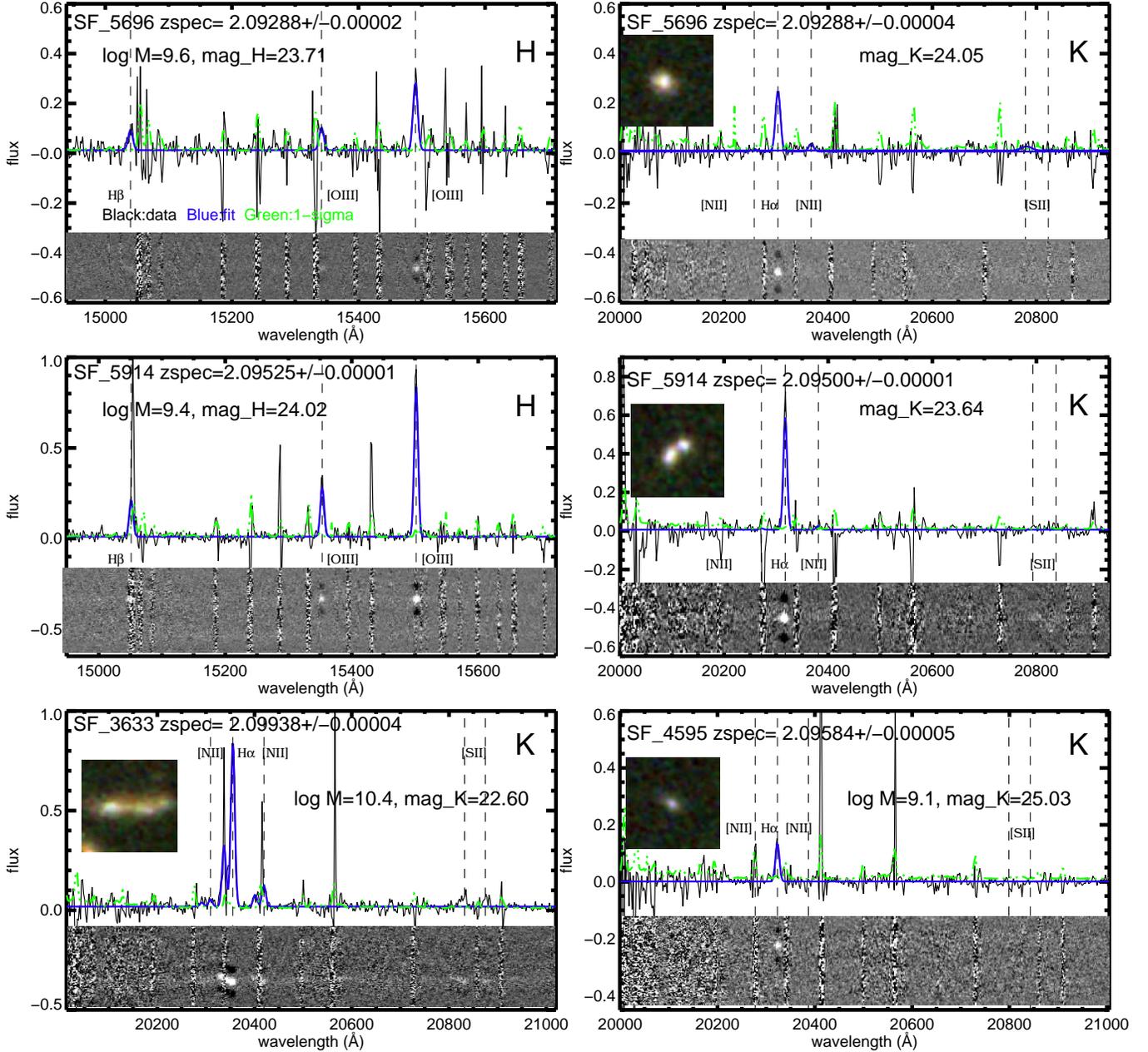}
\caption{Examples of the flux-calibrated MOSFIRE spectra for   cluster members.  We select 4 cluster galaxies of different brightness and 
and spectral quality that are representative of our  whole sample.  
Cluster members are observed primarily in the $K$ band; 1/4 of  the objects have  $H$ band observations.
On each panel, x-axis is the observed wavelength in $\AA$ and y-axis is  flux in unit of 10$^{-17}$ ergs~s$^{-1}$~cm$^{2}$~$\AA^{-1}$.
The observed 1-D spectra are presented in black and unsmoothed; the best-fit Gaussian line profiles   are superposed in blue;
the 1-$\sigma$ error spectra  are over-plotted in green.   
Spectroscopic redshifts from the Gaussian centroid fitting and associated 
statistical errors  are labeled. 
Vertical dashed lines show the expected positions of the strong emission lines at  spectroscopic redshifts. 
The photometric magnitude and stellar mass for each object are also marked.
Embedded are  2\arcsec$\times$2\arcsec three-color HST images (using the F814W, F125W and F160W filters) obtained from the publicly available CANDELS imaging \citep{Koekemoer11,Grogin11}.  
}
\label{fig:figzfourge}
\end{figure*}

\section{SPECTROSCOPIC OBSERVATIONS}

\subsection{MOSFIRE sample selection and observations}

We select spectroscopic targets based on the photometric redshifts in
ZFOURGE that were derived from imaging in deep near-infrared
medium-band filters \citep{Spitler12,Spitler14}.
The $z \sim 2$ galaxy cluster candidate was first discovered within
the COSMOS field in a single pointing of
$\sim11\arcmin\times11\arcmin$ targeted by ZFOURGE
\citep{Spitler12}.  The median uncertainties for the ZFOURGE
photometry is $\sim$ 0.05 dex \citep{Tomczak14}, sufficient to allow for efficient
cluster member candidates selection.

We obtained the spectroscopic data on MOSFIRE on the KECK 1 telescope
on Mauna Kea.  We conducted our observations on December 24-25, 2013
and February 10-13, 2014 with the aim of 1) securing as many redshifts
as possible in the field of the cluster candidate and 2) obtaining
high S/N spectra to study the physical properties (e.g.,
mass-metallicity relation, Kacprzak et al. in prep) of the cluster
members.  We configured 8 masks in the $K$-band filter covering
1.93-2.45 \micron\ (sensitive for detecting \ha\ and \nii\ lines at
$z\sim$ 2), and 2 masks in the $H$-band filter covering 1.46-1.81
\micron\ (sensitive for detecting \hb\ and \oiii\ lines at $z\sim$ 2).
We use a $0\farcs7$ slit width which yields a spectral resolution of  $R{=}3690$ in $K$ and $R{=}3620$ in $H$ band.

Taking advantage of the $6\farcm1\times6\farcm1$ MOSFIRE field of view
we targeted 224 objects in 6 pointings and secured redshifts for 180
objects.  The total on-source exposure time for the $K$-band masks is
$\sim$ 2 hours each. For the two $H$-band masks, the exposure is
5.3 and 3.2 hours respectively.  The observing conditions were
excellent for most of our masks, with seeing FWHM varying from $\simeq
0\farcs4$ to $\simeq 0\farcs7$.  An A0V type standard star was
observed in both the wide-slit mode and the narrow-science-slit
($0\farcs7$ slit width) mode before and after our science
targets.  The standard stars are used for telluric and flux
calibration.

\begin{figure}[!ht]
\centering
\includegraphics[trim = 2mm 7mm 6mm 2.mm, clip, width=6.4cm,angle=90]{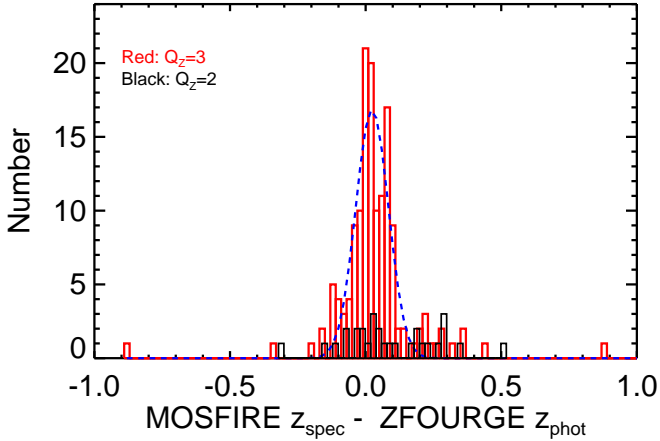}
\caption{Histogram of the  difference between our MOSFIRE spectroscopic redshifts and ZFOURGE photometric redshifts.  Binsize is 0.02. 
Objects with Q$_{\rm z}{=}3$ redshift  measurements (more than 2 emission lines identified) are shown in red and objects with Q$_{\rm z}{=}2$   redshift measurements (one emission line) are shown in black. 
Typical statistical errors for the MOSFIRE spectroscopic redshifts are 0.0001 whereas the median errors for  ZFOURGE photometric redshifts are 0.05.  The histogram can be well-fit by a Gaussian distribution
(blue dashed line).  The 1$\sigma$ scatter in the difference between the spectroscopic and photometric redshifts for Q$_{\rm z}{=}3$ objects is $\sigma$(Gaussian)/(1.+z({=}2)) $\sim$ 2 percent. 
}
  \label{fig:figphotoz}
\end{figure}

\subsection{Data Reduction and Redshift Measurements}

The raw MOSFIRE data were reduced using the publicly-available data
reduction pipeline (DRP) developed by the instrument team\footnote{See
\href{http://code.google.com/p/mosfire/}{http://code.google.com/p/mosfire/}}
available at the time.  The output of the MOSFIRE DRP were
background-subtracted, rectified and wavelength calibrated 2-D
spectra (see Figure~\ref{fig:figzfourge}).  All spectra were calibrated to vacuum wavelengths.  The
typical residual for the wavelength solution is   $\la$ 0.1 $\AA$.

Similar to the procedure used in \citet{Steidel14}, we develop our own
IDL routines to implement the telluric correction and flux calibration
based on the standard stars.    The 1-D
spectrum and its associated 1$\sigma$ error spectrum are extracted
using an aperture that corresponds to the FWHM of the spatial profile
of the well detected object (S/N $>$ 5). For objects that are too
faint to fit a Gaussian spatial profile, we use the FWHM of the
stellar profile on the same mask as the extraction aperture. 

Gaussian profiles were fit simultaneously to user-defined emission
lines, e.g. \ha\ and \nii, with the line center and velocity width
constrained to be the same within a given $K$-band or $H$-band.  Most
of our targets can be well-fitted by a single Gaussian component in
the spectral direction.  However, some of the galaxies in our sample
have good resolved velocity structures in the emission lines due to
great seeing (e.g, bottom left panel in Figure~\ref{fig:figzfourge}).  Those spectra require multiple component fitting and
will be presented in our future kinematic work of the sample. The
output of the code includes redshift, line flux, line width, and the
associated errors.  The statistical errors for each parameter are
estimated using the 1$\sigma$ error spectrum of the DRP, which we have
tested to represent the correct level of variation of the spectrum.

Examples of our reduced MOSFIRE spectra are presented in 
 Figure~\ref{fig:figzfourge}.  We show 4 cluster galaxies of different brightness and 
and spectral quality that are representative of our  whole sample.   The  $K$ band magnitude (AB) range of our cluster galaxies 
is $20.8 \le Ks \le 26.1$, with a median value of 23.86 (Nanayakkara et al., in preparation).  The faintest objects that we have detected emission lines (S/N $>$ 5) have  $Ks \sim 25$.

We flag the final redshifts in 3
categories based on the reliability of the redshift identification and measurements.

\begin{itemize}

\item For objects with at least 2 emission lines identified  at S/N $>5$, we flag them as  ``Q$_{\rm z}{=}3$", meaning the quality of the redshift are the highest and we are  confident that
the line identification and redshift measurement are correct.

\item For objects that show only 1 emission line with S/N $>5$, we
assign them as ``Q$_{\rm z}{=}2$" redshifts.  The general match of the
``Q$_{\rm z}{=}2$" object redshifts with their photometric redshifts suggests
that the single line identification is most likely to be correct (see
Figure~\ref{fig:figphotoz}).  The ``Q$_{\rm z}{=}2$" objects also show a spike at the cluster
redshift, further validating the ``Q$_{\rm z}{=}2$" redshifts
(Figure~\ref{fig:figdisz}).  The rms scatter between our spectroscopic
redshifts and the ZFOURGE photometric redshifts is about 5\%
(Figure~\ref{fig:figzfourge}).

\item For objects that have no obvious line detection (i.e., S/N $<
5$), we assign them as ``Q$_{\rm z}{=}1$" redshifts and do not include them in
the spec-z sample.

\end{itemize}

In summary, we identify 150 Q$_{\rm z}{=}3$ objects, 30 Q$_{\rm z}{=}2$ objects, and 44
Q$_{\rm z}{=}1$ objects ranging  from spectroscopic $z \sim$ 1.9 to 3.0.
The statistical errors determined from the fit to Gaussian centroids  for Q$_{\rm z}{=}2$ and Q$_{\rm z}{=}3$
redshifts are in the range of  $\Delta z \sim 0.0001-0.0002$ (median{=}0.00008).
To examine the systematic uncertainties, we compare the  redshifts of the Q$_{\rm z}{=}3$ objects (N $\sim$ 40) that have redundant observations with S/N $\ge 10$ in
both the $K$ and $H$ band.    The agreement between the redundantly detected redshifts 
is $\Delta z (\rm median) {=} 0.00005$ and  $\Delta z (\rm rms) {=} 0.00078$.  
We thus  quote  $\Delta z (\rm rms) {=} 0.00078/\sqrt{2}{=}0.00055$ as the total uncertainty of our redshift measurement which is 
contributed mostly from systematic uncertainties. 
At   $z{=}2.1$, this error  corresponds  to  a rest-frame velocity uncertainty of 
$\Delta {\rm v (rms)} {=} 53$ $\kms$ (spectral resolution $\sim$ 26 $\kms$).

\section{Results}

\subsection{Redshifts and Cluster Velocity Dispersion}


\begin{figure}[!ht]
\begin{center}
\includegraphics[trim = 5mm 0mm 10mm 3mm, clip, width=6.3cm,angle=90]{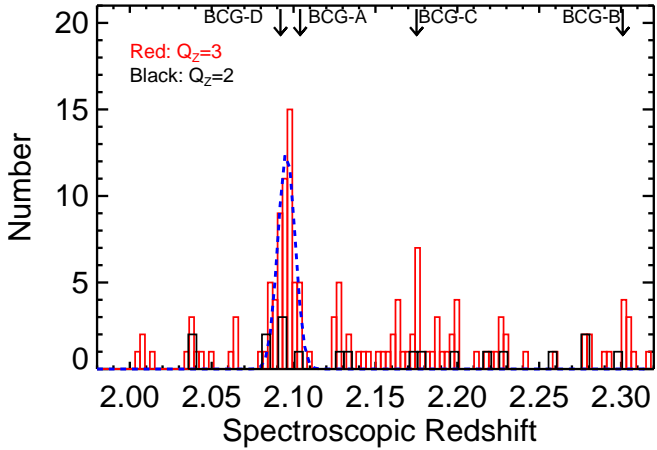}
\caption{Histogram for the redshift distribution  of galaxies in our sample that fall in the range of $2.0 \la z \la 2.3$. The binsize is 0.003.
Q$_{\rm z}{=}3$ objects are shown in red, Q$_{\rm z}{=}2$ in black.   A strong Gaussian-shaped spike is seen at redshift z$=$2.095.  
There are 57 galaxies that fall within the 3-sigma Gaussian width of the redshift peak.  We denote those 57 galaxies as  cluster members.
The spectroscopic redshifts for the brightest cluster galaxies (BCG-A, B, C, D) in the over density maps of \citet{Spitler12} are
marked with downward arrows. We have confirmed BCG-A and D to be the most massive red galaxies in the z$=$2.095 cluster, whereas BCG-B and C
are most likely associated with two background structures. 
}
\label{fig:figdisz}
\end{center}
\end{figure}

We show in Figure~\ref{fig:figdisz} the histogram of the spectroscopic
redshifts for Q$_{\rm z}{=}3$ (red) and Q$_{\rm z}{=}2$ objects (black).  
The redshift range of $2.0 < z < 2.3$ are used to exclude obvious interlopers.
A prominent spike at $z{=}2.095$ is clearly revealed.  The histogram
distribution around the spike can be well quantified by a Gaussian
profile (blue dashed line) with the center $z_{\rm c}$(Gaussian)${=}2.09578$ and
dispersion $\sigma_{\rm z}$ (Gaussian)${=}0.00544$ or in velocity
space $\sigma_{\rm v}$(Gaussian)${=}572$ km/s.  The mean of the
redshift distribution within identical redshift interval is $z_{\rm
c}$(mean)${=}2.09521$ and standard deviation $\sigma_{\rm z}$
(stdev)${=}0.00578$, skewness${=}-0.2318$, and kurtosis${=}-0.1753$ consistent with a Gaussian normal distribution.

We define $z_{\rm c}$ $\pm$ 3*$\sigma$z as the redshift range for the
cluster.  There are 57 galaxies that fall in this range, of which 52
are Q$_{\rm z}{=}3$ objects, and 5 are Q$_{\rm z}{=}2$ objects.  Whether the Q$_{\rm z}{=}2$
galaxies are included or not does not change our results.  

With MOSFIRE's spectral resolution and 57 confirmed members, we are
able to measure for the first time a robust cluster velocity
dispersion at $z\sim2$.  To calculate the cluster velocity dispersion
and errors, we bootstrap (with replacement) the 57 galaxies 30000 times and have: 
$z_{\rm c}$(boot)${=}2.09521 \pm 0.00076$ and dispersion $\sigma_{\rm z}$(boot)${=}0.00571 \pm 0.00053$ or in velocity units $\sigma_{\rm v1D}$(boot)${=}553 \pm 52$ km/s. To compare with previous sample sizes
(typically $\sim$10 members), we randomly select 10 galaxies from our
57 members and recalculate the bootstrapped (with replacement) velocity dispersion, we
obtain $\sigma_{\rm v1D}$(boot)${=}566 \pm169$ km/s, i.e. the
uncertainty in the velocity dispersion would be $\sim3$ times larger.

Because our spectroscopic catalog is biased towards star-forming galaxies, 
we are likely to have missed the quiescent/dusty galaxies or galaxies with faint emission lines below our detection limit (\ha\ 1$\sigma$ flux limit of our MOSFIRE survey is 3.2$\times$ $10^{-18}$ ergs~s$^{-1}$~cm$^{2}$; SFR$\sim$0.8 M$_\odot$ at $z{=}2.1$ without dust correction).  It has been shown that blue galaxies
in clusters have a larger velocity dispersion than red galaxies \citep[e.g.,][]{Carlberg97}.  
 Our velocity dispersion measurement could  be
slightly over-estimated due to this bias.  We defer the full analysis of this bias to future work.

\subsection{Spatial Distribution}

\begin{figure*}[!ht]
\begin{center}
\includegraphics[trim = 2mm 18mm 10mm 10mm, clip, width=13.4cm,angle=90]{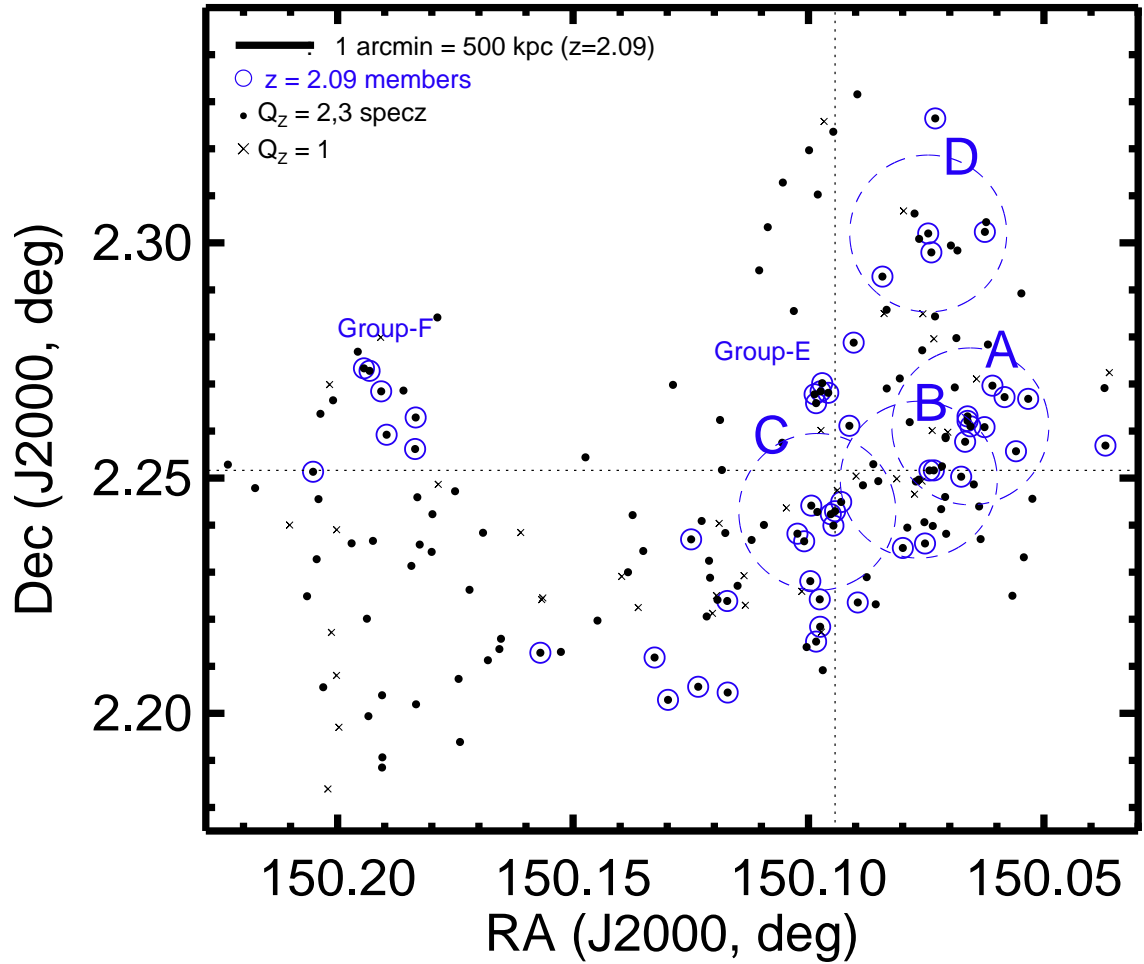}
\caption{MOSFIRE spectroscopic redshifts in the cluster candidate field.  
Small black dots show the all the 180 galaxies with reliable spectroscopic redshift identifications (``Q$_{\rm z}{=}3$" and ``Q$_{\rm z}{=}2$"  objects).
Black crosses show the 44 objects with no detections (``Q$_{\rm z}{=}1$" objects). 
Empty blue circles show the spatial distribution of the $z{=}2.095$ cluster members. 
The  dashed-line rings A, B, C, D denote the four peaks on the seventh nearest-neighbor surface density
maps as labeled in \citet{Spitler12}. We adopt the brightest galaxy as an overdensity's center. The coordinates for ABC centers are the same as \citet{Spitler12} 
and for D (10:00:17.739, +02:17:52.68, J2000) (Allen et al., in prep). 
The dotted lines marked  the median position of the cluster members  (10:00:22.646,+02:15:05.91).
The  rings have a radius of 1 arcmin which corresponds to a proper scale of 500 kpc at the cluster redshift of 2.09.
}
\label{fig:figspace}
\end{center}
\end{figure*}

\begin{figure}[!ht]
\begin{center}
\includegraphics[trim = 4mm 2mm 3mm 10mm, clip, width=6.8cm,angle=90]{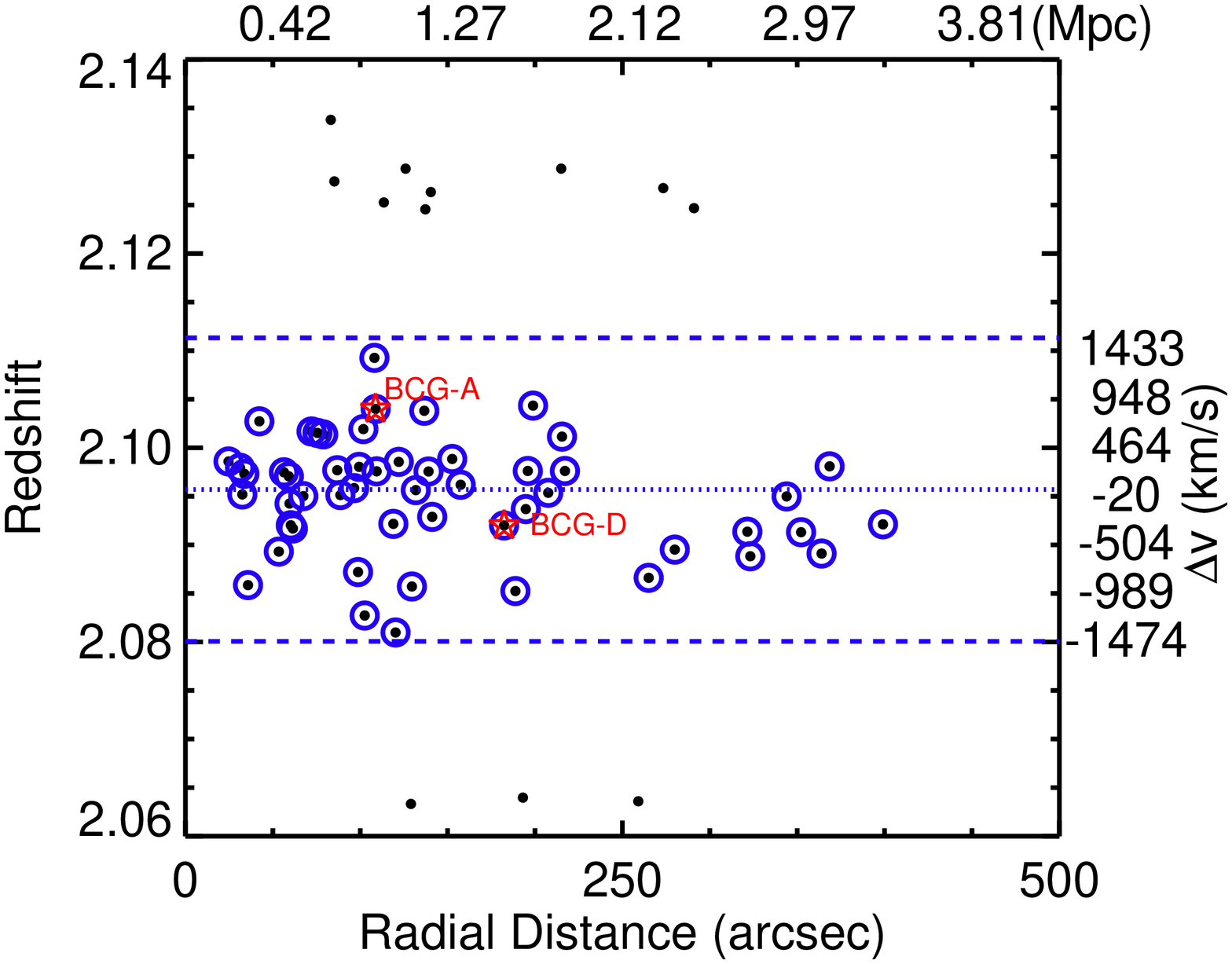}
\caption{Spectroscopic redshift (all Q$_{\rm z}{=}2,3$ objects, filled black dots) against radial distance to the ``center" of the cluster defined on the median position of the cluster members.
 The blue horizontal lines show the 
positions of the redshift peak (dotted) of $z=2.095$ and 3-sigma ranges (dashed).   35/57 of the 
members are concentrated within  1.3 Mpc  of the cluster  center.  The positions of the two massive red galaxies BCG-A and BCG-D are highlighted in red stars.
The velocities with respect to the cluster redshift $z=2.095$ are labeled on the y-axis on the right.
}
\label{fig:figrad}
\end{center}
\end{figure}

The spatial distribution of our MOSFIRE targets are presented in
Figure~\ref{fig:figspace}.  As described in \citet{Spitler12}, 3
strong overdensities (A, B, C) in this field are found by computing
surface density maps in narrow $\delta z{=}0.2$ redshift slices between
$z{=}1.5-3.5$ using the 7th nearest-neighbor metric \citep[e.g.,
][]{Papovich10,Gobat13}.  We also include another over-density region
D (Figure~\ref{fig:figspace}) using the same algorithm (Allen et al., in prep).  In each
over-density region, massive (M $>$ 10$^{11}$ M$_\odot$) galaxies are
selected as candidate brightest cluster galaxies (BCGs).  The positions of the
BCGs are taken as the overdensity's centers.  We also labeled in Figure~\ref{fig:figspace} Group E and F, which are 
groups of confirmed galaxies that are spatially concentrated and separated from the main structure.

Based on our MOSFIRE spectra, BCG-A and BCG-D are confirmed to be
quiescent galaxies that show only continua. Unfortunately, our
$K$-band and $H$-band observations do not cover obvious stellar
features for meaningful spectral template fitting.  We obtain the
spectroscopic redshifts for BCG-A (zspec{=}2.104) and BCG-D
(zspec{=}2.092) from \citet{Belli14}.  Our MOSFIRE spectra clearly
show that BCG-B and BCG-C are star-forming emission-line galaxies that
lie at redshift $z{=}2.3010 \pm 0.0001$ and 2.1750 $\pm$ 0.0001
respectively.  The photometric redshift of BCG-B and C is $2.15^{+0.05}_{-0.06}$
and
$2.19^{+0.04}_{-0.03}$, both are under-estimated, especially for BCG-B.

The number of members with projected radius of 500 kpc for  the original \citet{Spitler12} ABC,  and D overdensities are 12, 5, 8, and 4 respectively (Figure~\ref{fig:figspace}), though
 we note that what we called ``BCGs" B,C are behind the main structure indicating the dangers of studying membership based solely on photometric data.

The 57 cluster members cover a total projected spatial length of
$\sim$ 3.7$\times$5 Mpc$^2$ ($\sim$ 7.4 $\times$ 10 Mpc$^2$  comoving).  Taking the median position of the 57 cluster members (dotted
lines in Figure~\ref{fig:figspace}) as the cluster center, we show the
radial distance of members from this defined cluster center in
Figure~\ref{fig:figrad}.  Note there are 35 members that fall within
the 1.3 Mpc projected radius over the multiple over-density peaks.

\section{Comparison with simulations}\label{highz}

To help us understand what our observed structure at $z{=}2.095$  should evolve into at $z{=}0$, we employ the $2160^3$ particle Gpc-volume (particle mass $m_p{=}1.1\times10^{10} {\rm M_\odot}$) GiggleZ-main simulation \citep{Poole14}.  We have computed 1D velocity dispersions for all the friends-of-friends (FoF) structures of the simulation at $z{=}0$ and $z{=}2.2$ (the closest snapshot to our observed redshift) using substructures exceeding $\rm{M_{vir}}{=}4.3\times10^{11} {\rm M_\odot}$.  Merger trees were used to determine what each $z{=}$2.2 FoF structure evolves into at $z{=}0$.  We find that systems with velocity dispersions in the range $\sigma_{\rm 1D}{=}552\pm52$ km/s at $z{=}2.2$ have virial masses in the range ${\rm M_{vir}}{=}10^{13.5\pm0.2} {\rm M_\odot}$ and that they evolve into systems with ${\rm M_{vir}}{=}10^{14.4\pm0.3} {\rm M_\odot}$ and $\sigma_{\rm1D}{=}680^{+73}_{-110}$ km/s (all ranges are $68\%$ confidence), in agreement with a Virgo-like cluster  \citep{Vaucouleurs61}.
 299 such systems are found in the simulation suggesting an incidence of one per 2.5 square degrees over the redshift range $z{=}2.0$ to $2.3$.
This corresponds to a $\sim$ 4\% occurrence of such a cluster in the original ZFOURGE survey area of 0.1 deg$^{2}$. 

These results are in good agreement with the $z{=}2$ $\sigma_{sub}{-}{\rm M_{vir}}$ relation of \citet{Munari13} and 
with the results of \citet{Chiang13} whose simulations indicate that a $10^{13.5} {\rm M_\odot}$ system should evolve to a mass of ${\sim}10^{14.5} {\rm M_\odot}$ at $z{=}0$.

\section{Conclusions}\label{discussion}

We carry out MOSFIRE spectroscopic observations in the $z\sim 2$
galaxy cluster candidate with a red-sequence that was first discovered from the Magellan/\fourstar\
Galaxy Evolution Survey (ZFOURGE) \citep{Spitler12}.  This galaxy
cluster was identified using rest-frame optical and near-infrared
imaging and is thus an important link between the UV-selected systems
at this epoch \citep[e.g.,][]{Steidel05,Digby-North10} and massive
clusters at lower-redshift \citep[e.g.,][]{Gal04}.  Our successful
spectroscopic campaign confirms the accuracy of the photometric
redshifts derived from ZFOURGE's deep medium-bandwidth photometry.

By combining MOSFIRE's spectral capabilities  with
our efficient selection of $z\sim2$ targets, we are able to identify
cluster members and accurately measure the cluster's kinematics.  We
measure spectral redshifts for 180 objects and identify 57 cluster
members that have a mean redshift of $z{=}2.095$.  The redshifts for
cluster members are determined primarily from \ha\ and \nii\ emission,
and the cluster velocity dispersion is $\sigma_{\rm v1D}$ {=} 552 $\pm$
52 km/s.  Most of the cluster galaxies (35) lie within a region with a
projected radius of 1.3 Mpc.

This is the first study of a galaxy cluster at $z \ge 2.0$ with the
combination of spectral resolution ($\sim$ 26 km/s) and the number of
confirmed members ($>$ 50) needed to study cluster kinematics robustly
and map members over a large field of view (12 $\times$ 12
arcmin$^2$).  Our accurate velocity dispersion measurement of this
clustering structure allows us to use simulations to trace the cluster's
likely evolution to $z{=}0$.  Our simulation results show that
the ZFOURGE cluster at $z=2.095$ should evolve into a Virgo-like system locally
with ${\rm M_{vir}}{=}10^{14.4\pm0.3} {\rm M_\odot}$.

Our results show that galaxy clusters at $z\sim2$ can now be studied
in the same detailed manner as clusters at $z\la 1$.  However, unlike
galaxies in massive clusters at $z\sim0$, the ZFOURGE cluster members
show a wealth of \ha\ emission and other signs of star formation
activity.  Our next work will report the mass-metallicity relation,
ionization parameter evolution and other physical properties of the
ZFOURGE cluster at $z=2.095$.

\acknowledgments 
We would like to thank the referee for an excellent report and comments that have improved this paper.
We thank Pierluigi Cerulo for useful comments.
KG, LS, TN, acknowledges funding from a Australian Research Council
(ARC) Discovery Program (DP) grant DP1094370 and Access to Major
Research Facilities Program which is supported by the Commonwealth of
Australia under the International Science Linkages program.
L.K. acknowledges a NSF Early CAREER Award AST
0748559 and an ARC Future Fellowship award FT110101052.  
GP acknowledges support from the ARC Laureate Fellowship of Stuart Wyithe.
Observations were supported by Swinburne Keck programs 2013B\_W160M
and 2014A\_W168M.  Part of this
work was supported by a NASA Keck PI Data Award, administered by the
NASA Exoplanet Science Institute.  The authors wish to recognize and
acknowledge the very significant cultural role and reverence that the
summit of Mauna Kea has always had within the indigenous Hawaiian
community. We are most fortunate to have the opportunity to conduct
observations from this mountain.


\end{document}